\documentclass[12pt]{article}

\begin{document}
\begin{center}
\Large{\bf Classical Capacity of A Quantum Multiple Access
Channel}\footnote{ The project supported by National Natural
Science Foundation of
China}\\
\bigskip\bigskip
 \large{Minxin Huang$^{\dag}$\footnote{Email:minxin@ustc.edu}, Yongde Zhang$^{\S\ddag}$, Guang
Hou$^{\ddag}$}\\
\bigskip\bigskip
\normalsize{
$^{\dag}${ Special Class for Gifted Young,}\\
{ University of Science and Technology of China,}\\
{ Hefei, 230026, P.R. China.}\\
$^{\S}${ CCAST (World Laboratory), P.O. Box 8730,}\\
{ Beijing 100080, P.R. China.}\\
$^{\ddag}${ Department of Modern Physics,}\\
{ University of Science and Technology of China.}\\
{ Hefei, 230027, P.R. China.}}
\end{center}

\begin{abstract}
We consider the transmission of classical information over a quantum channel
by two senders. The channel capacity region is shown to be a convex hull
bound by the {\em Von Neumann} entropy and the conditional {\em Von Neumann}
entropy. We discuss some possible applications of our result. We also show
that our scheme allows a reasonable distribution of channel capacity over
two senders.
\end{abstract}


\vskip 1.0cm

\section{Introduction}

In quantum information theory, a basic question often presented is how
efficiently can one transmit classical information over a quantum channel.$^{%
\cite{jpG,lbL,asH}}$ Because of the non-orthogonality of quantum states, the
channel capacity is different from that of the classical channel discussed
in $^{\cite{ceS,tmCjaT}}$ . Recently, this question has attracted new
attention due to the rapid progress in quantum information theory. In
particular, a theorem is established that the maximum attainable rate of
asymptotically error free transmission of classical information over a
quantum channel is precisely the {\em Holevo} bound.$^{\cite
{pHrJbSmWwkW,bSmW,H,jP}}$

In this paper, we consider the transmission of classical information by two
senders to a common receiver. The scheme can be viewed as a quantum multiple
access channel, which is the quantum analogy of classical multiple access
channel.$^{\cite{tmCjaT}}$ Suppose two senders, {\em Alice} and {\em Bob},
are given an ensemble of normal letter states $\left| \Psi _{\alpha \beta
}\right\rangle $, where $\alpha $, $\beta $ can be drawn from two alphabet
sets: ${\cal H}_A=\left\{ \alpha \right\} $ and ${\cal H}_B=\left\{ \beta
\right\} $, while {\em Alice} is allowed to choose the letter $\alpha $,
{\em Bob} is allowed to choose the letter $\beta $. Then the letter state is
sent to the receiver, {\em Charlie}, who subjects it to a measurement to
determine which letters {\em Alice} and {\em Bob} have chosen. We should
note that the sets $\left\{ \alpha \right\} ,\left\{ \beta \right\} $ are
purely classical alphabets in the hands of Alice and Bob, and $\left| \Psi
_{\alpha \beta }\right\rangle $ does not imply a tensor product {\em Hilbert}
space. For example, if $Alice$ uses the two-symbol alphabet $\left\{
A,B\right\} $, and Bob uses $\left\{ C,D\right\} $, then it is still
possible that the signal states live in a two-dimensional $Hilbert$ space ,
that is $\left| \Psi _{AC}\right\rangle =\left| 0\right\rangle $, $\left|
\Psi _{AD}\right\rangle =\left| 1\right\rangle $, $\left| \Psi
_{BC}\right\rangle =\frac 1{\sqrt{2}}(\left| 0\right\rangle +\left|
1\right\rangle )$, $\left| \Psi _{BD}\right\rangle =\frac 1{\sqrt{2}}(\left|
0\right\rangle -\left| 1\right\rangle )$. So entanglement is not at issue
here at all. But we will see it coming up in the application in Sec 6.

Suppose $p_\alpha q_\beta $ is a product probability distribution of the
letter state $\left| \Psi _{\alpha \beta }\right\rangle $, then denote the
density matrix by

\[
\begin{array}{lll}
\rho & = & \sum\limits_{\alpha ,\beta }p_\alpha q_\beta \left| \Psi _{\alpha
\beta }\right\rangle \left\langle \Psi _{\alpha \beta }\right| \\
\rho _\alpha & = & \sum\limits_\beta q_\beta \left| \Psi _{\alpha \beta
}\right\rangle \left\langle \Psi _{\alpha \beta }\right| \\
\rho _\beta & = & \sum\limits_\alpha p_\alpha \left| \Psi _{\alpha \beta
}\right\rangle \left\langle \Psi _{\alpha \beta }\right|
\end{array}
\]

We denote the conditional {\em Von Neumann} entropy by

\begin{equation}
\begin{array}{lll}
H_A & = & \sum\limits_\beta q_\beta H\left( \rho _\beta \right) \\
H_B & = & \sum\limits_\alpha p_\alpha H\left( \rho _\alpha \right)
\end{array}
\label{eq01}
\end{equation}
We can see $H_A,H_B\leq H\left( \rho \right) $ from the concavity of {\em %
Von Neumann} entropy$^{\cite{aW}}$. A code $\left( \left( M,N\right)
,l\right) $ is defined to consist of $M$ $\alpha $-letter sequences and $N$ $%
\beta $-letter sequences of length $l$, and together they form $MN$ code
words $\left\{ \left| S_{ij}\right\rangle :i=1,2,\cdots ,M;j=1,2,\cdots
,N\right\} $, with each $\alpha $-letter sequence and $\beta $-letter
sequence combined. We use the above example to make things clear . Suppose $%
l=2$, and $Alice$ choose $2$ $\alpha $-letter sequences $\left\{
AA,AB\right\} $, Bob choose $\left\{ CD,DD\right\} $, then $M=N=2$, and we
see that $\left| S_{11}\right\rangle =\left| \Psi _{AC}\right\rangle \otimes
\left| \Psi _{AD}\right\rangle $, $\left| S_{12}\right\rangle =\left| \Psi
_{AD}\right\rangle \otimes \left| \Psi _{AD}\right\rangle $, $\left|
S_{21}\right\rangle =\left| \Psi _{AC}\right\rangle \otimes \left| \Psi
_{BD}\right\rangle $, and $\left| S_{22}\right\rangle =\left| \Psi
_{AD}\right\rangle \otimes \left| \Psi _{BD}\right\rangle $.

We assume the $MN$ code words have the same probability $\frac 1{MN}$. A
rate $\left( R_1,R_2\right) $ is said to be achievable if there exists a
sequence of $\left( \left( 2^{nR_1},2^{nR_2}\right) ,n\right) $ codes for
which {\em Charlie} can decode the message by a measurement with error
probability $P_E\rightarrow 0$ when $n$ tends to infinity. The capacity
region of the multiple access channel is the closure of the set of
achievable $\left( R_1,R_2\right) $ rate pairs. The following theorem is the
main result of this paper.

{\bf Theorem} The capacity region of a quantum multiple access channel is
the closure of the convex hull of all $\left( R_1,R_2\right) $ satisfying

\begin{equation}
R_1<H_A,R_2<H_B,R_1+R_2<H\left( \rho \right)  \label{eq02}
\end{equation}
for some product distribution $p_\alpha q_\beta $ on ${\cal H}_A\times {\cal %
H}_B$.

Thus $H_A$, $H_B$ are analogs of classical conditional mutual information.$^{%
\cite{tmCjaT}}$ Our result provides an information-theoretical
interpretation of the conditional {\em Von Neumann} entropy.

We should note that the problem addressed here have been considered and
solved in previous papers. It was first raised by Allahverdyan and Saakian%
\cite{aea}, who essentially discussed the converse theorem in this paper.
Then Winter gave general formulation and solution for many users and noisy
channels\cite{aw}. See also Winter's Ph.D. Thesis\cite{awi}. However,
Winter's proof is highly abstract, and he himself spoke of desirability of a
more direct proof. This is what is done in this paper, for the particular
case of two users and noiseless channel.

The paper is organized as follows. In {\rm Section 2}, we describe some
useful properties which are necessary to the proof. In {\rm Section 3} and
{\rm 4}, we prove the achievability of this theorem. In {\rm Section 5}, we
prove the converse theorem. Finally, in {\rm Section 6}, we discuss some
applications of the theorem.

\section{Some useful properties}

We first prove a lemma about conditional {\em Von Neumann} entropy.

{\bf Lemma 1} $p_\alpha q_\beta $ is a fixed product distribution, then

\begin{equation}
H_A+H_B\geq H\left( \rho \right)  \label{eq03}
\end{equation}

{\bf [Proof]} This lemma can be proved by the strong subadditivity of {\em %
Von Neumann} entropy$^{\cite{aW}}$. If denote

\[
\rho ^{RST}=\sum\limits_{\alpha ,\beta }p_\alpha q_\beta \left| \Psi
_{\alpha \beta }\right\rangle \left\langle \Psi _{\alpha \beta }\right|
\otimes \left| i_\alpha \right\rangle \left\langle i_\alpha \right| \otimes
\left| j_\beta \right\rangle \left\langle j_\beta \right|
\]
here $\left| i_\alpha \right\rangle $, $\left| j_\beta \right\rangle $ are
orthogonal states in {\em Hilbert} space $S$ and $T$, then

\[
\begin{array}{lll}
\rho ^R & = & \rho \\
\rho ^{RS} & = & \sum\limits_{\alpha ,\beta }p_\alpha q_\beta \left| \Psi
_{\alpha \beta }\right\rangle \left\langle \Psi _{\alpha \beta }\right|
\otimes \left| i_\alpha \right\rangle \left\langle i_\alpha \right| \\
\rho ^{RT} & = & \sum\limits_{\alpha ,\beta }p_\alpha q_\beta \left| \Psi
_{\alpha \beta }\right\rangle \left\langle \Psi _{\alpha \beta }\right|
\otimes \left| j_\beta \right\rangle \left\langle j_\beta \right|
\end{array}
\]
we use the strong subadditivity of {\em Von Neumann} entropy:

\[
H\left( \rho ^{RST}\right) +H\left( \rho \right) \leq H\left( \rho
^{RS}\right) +H\left( \rho ^{RT}\right)
\]
It can be easily seen that

\[
\begin{array}{lll}
H\left( \rho ^{RST}\right) & = & H\left( p_\alpha q_\beta \right) =H\left(
p_\alpha \right) +H\left( q_\beta \right) \\
H\left( \rho ^{RS}\right) & = & H\left( p_\alpha \right) +H_B \\
H\left( \rho ^{RT}\right) & = & H\left( q_\beta \right) +H_A
\end{array}
\]
where $H\left( p_\alpha q_\beta \right) $, $H\left( p_\alpha \right) $ and $%
H\left( q_\beta \right) $ are {\em Shannon} entropies. Thus, we have

\[
H_A+H_B\geq H\left( \rho \right)
\]

Another property of the quantum multiple access channel is its convexity,
{\it i.e.}, if $\left( R_1,R_2\right) $ and $\left( R_1^{^{\prime
}},R_2^{^{\prime }}\right) $ are achievable rates, then $\left( \lambda
R_1+\left( 1-\lambda \right) R_1^{^{\prime }},\lambda R_2+\left( 1-\lambda
\right) R_2^{^{\prime }}\right) $ is also achievable, for any $0\leq \lambda
\leq 1$. The idea is the time sharing scheme. Given two sequences of codes
at different rates $\left( R_1,R_2\right) $ and $\left( R_1^{^{\prime
}},R_2^{^{\prime }}\right) $, we can construct a third code book at the rate
of $\lambda \left( R_1,R_2\right) +\left( 1-\lambda \right) \left(
R_1^{^{\prime }},R_2^{^{\prime }}\right) $ by using the first code book for
the first $\lambda n$ symbols and the second for the remaining $\left(
1-\lambda \right) n$ symbols. Since the overall probability of error is less
than the sum of the probability of error for each of the segments, the
probability of error of the new code approaches zero, the rate is achievable.

For a fixed product distribution $p_\alpha q_\beta $, since $H_A+H_B\geq
H\left( \rho \right) \geq H_A$ or $H_B$, according to the convexity of
capacity region, we need only to prove $\left( H\left( \rho \right)
-H_B,H_B\right) $ and $\left( H_A,H\left( \rho \right) -H_A\right) $ are
achievable rates in order to prove all rate pair satisfying Eq.(\ref{eq02})
is achievable. We will do this in {\rm Section 3} and {\rm 4}.

There is a useful inequality in bounding the decoding error. Suppose $%
\left\langle 1|1\right\rangle \leq 1$ and $\left\langle 2|2\right\rangle
\leq 1$ ,then

\[
\begin{array}{lll}
\left| \left\langle 1|3\right\rangle \right| & \leq & \left| \left\langle
2|3\right\rangle \right| +\left| \left( \left\langle 1\right| -\left\langle
2\right| \right) \left| 3\right\rangle \right| \\
& \leq & \left| \left\langle 2|3\right\rangle \right| +\sqrt{\left\langle
3|3\right\rangle }\sqrt{\left\langle 1|1\right\rangle +\left\langle
2|2\right\rangle -\left\langle 1|2\right\rangle -\left\langle
2|1\right\rangle } \\
& \leq & \left| \left\langle 2|3\right\rangle \right| +\sqrt{\left\langle
3|3\right\rangle }\sqrt{2-\left\langle 1|2\right\rangle -\left\langle
2|1\right\rangle }
\end{array}
\]
therefore

\begin{equation}
\left| \left\langle 2|3\right\rangle \right| \geq \left| \left\langle
1|3\right\rangle \right| -\sqrt{\left\langle 3|3\right\rangle }\sqrt{%
2-\left\langle 1|2\right\rangle -\left\langle 2|1\right\rangle }
\label{eq04}
\end{equation}
This inequality implies that if $\left\langle 1|3\right\rangle $ and $%
\left\langle 1|2\right\rangle $ are close to unity, then $\left\langle
2|3\right\rangle $ is also close to unity.

\section{Compound measurement}

As noted in previous sections, we only need to prove $\left( H\left( \rho
\right) -H_B,H_B\right) $ is achievable rate. We use latin character $a$, $b$%
, ... to index {\em Alice}'s strings, and $a^{^{\prime }}$, $b^{^{\prime }}$%
, ... to index {\em Bob}'s. Suppose the string length is $L$. Denote

\[
\rho _a=\sum\limits_{\textstyle{all strings }a^{^{\prime
}}}P_{a^{^{\prime }}}\left| S_{aa^{^{\prime }}}\right\rangle
\left\langle S_{aa^{^{\prime }}}\right|
\]
Here $P_{a^{^{\prime }}\textstyle{ }}$ means the product
probability (For
example, if $a^{^{\prime }}=CD$, then $P_{a^{^{\prime }}}=q_Cq_{D\textstyle{.}}$%
). The sum $a^{^{\prime }}$ is over all possible strings, so $\rho _a$ is a
product state. For example, if $a=AB$, then $\rho _a=\rho _A\otimes \rho _B$.

$\rho _a$ has a complete orthonormal set of eigenstates, which we denote as $%
\left| t_{ak}\right\rangle $, and a corresponding set of eigenvalues $%
p_{k|a} $. Let $\varepsilon ,\delta >0$. Then we can find a length $L$ long
enough to enforce some typicality conditions. Noticing that the quantity $%
H\left( \rho \right) -H_B$ is the {\em Holevo} information of the ensemble $%
\left\{ p_\alpha ,\rho _\alpha \right\} $, it was proved in Ref.\cite{bSmW}
that {\em Alice} can choose $M=2^{L\left( H\left( \rho \right) -H_B-\delta
\right) }$ strings, so that the decoder can distinguish the eigenstates of
the mixed states $\rho _a$ by a {\bf POVM}(Positive Operator Valued
Measure). Suppose $\left| \tilde u_{ak}\right\rangle \left\langle \tilde u%
_{ak}\right| $ are the elements of the decoding {\bf POVM}. Then for every
string $a$ (in the $M$ strings), the probability of right guess is

\begin{equation}
\sum\limits_{k} p_{k|a}\left| \left\langle \tilde u_{ak}|t_{ak}\right\rangle
\right| ^2>1-\varepsilon  \label{eq05}
\end{equation}

Denote $\Pi _a$ as the projection onto the subspace of vectors $\left|
t_{ak}\right\rangle $ satisfying

\begin{equation}
2^{-L\left( H_B+\delta \right) }<p_{k|a}<2^{-L\left( H_B-\delta \right) }
\label{eq06}
\end{equation}
Because the {\bf POVM} element $\left| \tilde u_{ak}\right\rangle =0$ when $%
p_{k|a}$ doesn't satisfy Eq.(\ref{eq06}) (as noted in Ref.\cite{bSmW}), and $%
\left\langle \tilde u_{ak}|\tilde u_{ak}\right\rangle \leq 1$, we have

\begin{equation}
tr\left( \Pi _a\rho _a\Pi _a\right) =\sum\limits_kp_{k|a}\geq
\sum\limits_kp_{k|a}\left| \left\langle \tilde u_{ak}|t_{ak}\right\rangle
\right| ^2>1-\varepsilon  \label{eq07}
\end{equation}

\begin{equation}
tr\left( \Pi _a\rho _a^2\Pi _a\right) \leq 2^{-L\left( H_B-3\delta \right) }
\label{eq08}
\end{equation}
We choose the $M$ strings described above as {\em Alice}'s signal strings
and we will use random code to select {\em Bob}'s signal strings.

The decoding process includes two measurements, the first can decode {\em %
Alice}'s signal and the second can decode {\em Bob}'s signal. Denote

\[
A_a=\sum\limits_k\left| t_{ak}\right\rangle \left\langle \tilde u%
_{ak}\right|
\]
Since $\sum\limits_aA_a^{\dagger }A_a=\sum\limits_{ak}\left| \tilde u%
_{ak}\right\rangle \left\langle \tilde u_{ak}\right| $, $A_a$ is a decoding
{\bf POVM} element, which will be the decoder's first measurement. Suppose
the result of the first measurement is string $a$, then the decoder's second
measurement will be the so called ''{\bf pretty good measurement}'' used in
Ref.\cite{pHrJbSmWwkW} to distinguish $N$ states $\left\{ \Pi _a\left|
S_{aa^{^{\prime }}}\right\rangle \right\} $ in order to determine $%
a^{^{\prime }}$. Suppose $\left| \tilde \eta _{a^{^{\prime
}}|a}\right\rangle \left\langle \tilde \eta _{a^{^{\prime }}|a}\right| $ is
the element of the {\bf POVM}. Together these two measurements form a
compound measurement. The probability of error is

\[
P_E=1-\frac 1{MN}\sum\limits_{aa^{^{\prime }}} \left| \left\langle \tilde
\eta _{a^{^{\prime }}|a}\right| A_a\left| S_{aa^{^{\prime }}}\right\rangle
\right| ^2
\]
we denote

\[
P_{Ea}=1-\frac 1N\sum\limits_{a^{^{\prime }}} \left| \left\langle \tilde \eta
_{a^{^{\prime }}|a}\right| A_a\left| S_{aa^{^{\prime }}}\right\rangle
\right| ^2
\]
then

\[
P_E=\frac 1MP_{Ea}
\]

Denote the random code average by ''$\left\langle {}\right\rangle _c$''. The
random code is averaged over {\em Bob}'s codes. We will prove $\left\langle
P_{Ea}\right\rangle _c<8\varepsilon $ for every string $a$.

\section{{\em Bob}'s random codes}

Using the {\em Schwarz} inequality and the inequality (\ref{eq04}), we have

\[
\begin{array}{lll}
\sqrt{1-P_{Ea}} & = & \left( \frac 1N\sum\limits_{a^{^{\prime }}} \left|
\left\langle \tilde \eta _{a^{^{\prime }}|a}\right| A_a\left|
S_{aa^{^{\prime }}}\right\rangle \right| ^2\right) ^{\frac 12} \\
& \geq & \frac 1N\sum\limits_{a^{^{\prime }}} \left| \left\langle \tilde \eta
_{a^{^{\prime }}|a}\right| A_a\left| S_{aa^{^{\prime }}}\right\rangle \right|
\\
& \geq & \frac 1N\sum\limits_{a^{^{\prime }}} \left| \left\langle \tilde \eta
_{a^{^{\prime }}|a}|S_{aa^{^{\prime }}}\right\rangle \right| \\
&  & -\frac 1N\sum\limits_{a^{^{\prime }}} \sqrt{\left\langle \tilde \eta
_{a^{^{\prime }}|a}|\tilde \eta _{a^{^{\prime }}|a}\right\rangle }\left(
2-\left\langle S_{aa^{^{\prime }}}\right| A_a\left| S_{aa^{^{\prime
}}}\right\rangle -\left\langle S_{aa^{^{\prime }}}\right| A_a^{\dagger
}\left| S_{aa^{^{\prime }}}\right\rangle \right) ^{\frac 12} \\
& \equiv & \Omega _1-\Omega _2
\end{array}
\]
here $\Omega _1=\frac 1N\sum\limits_{a^{^{\prime }}} \left| \left\langle
\tilde \eta _{a^{^{\prime }}|a}|S_{aa^{^{\prime }}}\right\rangle \right| $,\\%
$\Omega _2=\frac 1N\sum\limits_{a^{^{\prime }}} \sqrt{\left\langle \tilde
\eta _{a^{^{\prime }}|a}|\tilde \eta _{a^{^{\prime }}|a}\right\rangle }%
\left( 2-\left\langle S_{aa^{^{\prime }}}\right| A_a\left| S_{aa^{^{\prime
}}}\right\rangle -\left\langle S_{aa^{^{\prime }}}\right| A_a^{\dagger
}\left| S_{aa^{^{\prime }}}\right\rangle \right) ^{\frac 12}$.

According to {\em Schwarz} inequality,

\[
\begin{array}{lll}
\sqrt{\left\langle 1-P_{Ea}\right\rangle _c} & \geq & \left\langle \sqrt{%
1-P_{Ea}}\right\rangle _c \\
& \geq & \left\langle \Omega _1\right\rangle _c-\left\langle \Omega
_2\right\rangle _c
\end{array}
\]

We deal with $\left\langle \Omega _1\right\rangle _c$ and $\left\langle
\Omega _2\right\rangle _c$ respectively. The calculation of $\left\langle
\Omega _1\right\rangle _c$ is the same as in Ref.\cite{pHrJbSmWwkW}, it was
proved by using Eqs.(\ref{eq06},\ref{eq07},\ref{eq08}) that

\[
\left\langle \Omega _1\right\rangle _c\geq 1-\varepsilon -N\cdot 2^{-L\left(
H_B-3\delta \right) }
\]

Next, we examine term $\Omega _2$. First, we notice that $\left| \tilde \eta
_{a^{^{\prime }}|a}\right\rangle \left\langle \tilde \eta _{a^{^{\prime
}}|a}\right| $ is the {\bf POVM} element, so $\left\langle \tilde \eta
_{a^{^{\prime }}|a}|\tilde \eta _{a^{^{\prime }}|a}\right\rangle \leq 1$.
Then we have

\[
\begin{array}{cl}
\Omega _2 & \leq \left( \frac 1N\sum\limits_{a^{^{\prime }}} \left\langle
\tilde \eta _{a^{^{\prime }}|a}|\tilde \eta _{a^{^{\prime }}|a}\right\rangle
\right) \cdot \frac 1N\sum\limits_{a^{^{\prime }}} \left( 2-\left\langle
S_{aa^{^{\prime }}}\right| A_a\left| S_{aa^{^{\prime }}}\right\rangle
-\left\langle S_{aa^{^{\prime }}}\right| A_a^{\dagger }\left|
S_{aa^{^{\prime }}}\right\rangle \right) \\
& \leq 2-\frac 1N\sum\limits_{a^{^{\prime }}} \left( \left\langle
S_{aa^{^{\prime }}}\right| A_a+A_a^{\dagger }\left| S_{aa^{^{\prime
}}}\right\rangle \right)
\end{array}
\]

Averaged over {\em Bob}'s code, then

\[
\begin{array}{cl}
\left\langle \Omega _2\right\rangle _c & \leq 2-Tr\left[ \left(
A_a+A_a^{\dagger }\right) \rho _a\right] \\
& =2-2\sum\limits_kp_{k|a}\left| \left\langle \tilde u_{ak}|t_{ak}\right%
\rangle \right| \\
& \leq 2\left( 1-\sum\limits_kp_{k|a}\left| \left\langle \tilde u%
_{ak}|t_{ak}\right\rangle \right| ^2\right) \\
& <2\varepsilon
\end{array}
\]

So $\sqrt{\left\langle 1-P_{E_a}\right\rangle _c}>1-3\varepsilon
-N2^{-L\left( H_B-3\delta \right) }$, then we choose $N=2^{L\left(
H_B-4\delta \right) }$. When $L$ is large, we have $\sqrt{\left\langle
1-P_{E_a}\right\rangle _c}>1-4\varepsilon $, so $\left\langle
P_{E_a}\right\rangle _c<8\varepsilon $. And therefore

\[
\left\langle P_E\right\rangle _c=\frac 1M\sum\limits_{a} \left\langle
P_{Ea}\right\rangle _c<8\varepsilon
\]

The average probability of error is small, so {\em Bob} can find a
particular code for which $P_E<8\varepsilon $, thus complete the proof of
the achievability of the theorem.

\section{About the converse theorem}

Denote ${\cal E}$ as the closure of the convex hull of all $\left(
R_1,R_2\right) $ satisfying Eq.(\ref{eq02}). Suppose {\em Alice} and {\em Bob%
} can send information to {\em Charlie} at the rate of $\left(
R_1,R_2\right) $, then we shall prove that $\left( R_1,R_2\right) \in {\cal E%
}$. This is the converse of the theorem. Consider a $\left( \left(
2^{lR_1},2^{lR_2}\right) ,l\right) $ code in which the code words are $%
\left| S_{aa^{^{\prime }}}\right\rangle $. Suppose {\em Charlie} can decode
the signals asymptotically error free, then when $l$ is sufficiently large
we have the inequality

\begin{equation}
\begin{array}{cll}
R_1+R_2 & \leq & \frac 1lI\left( \textstyle{{ Charlie}}:\textstyle{{ Alice }},%
\textstyle{{ Bob}}\right) \\
R_1 & \leq & \frac 1lI\left( \textstyle{{ Charlie}}:\textstyle{{ Alice }}|\textstyle{%
{ Bob}}\right) \\
R_2 & \leq & \frac 1lI\left( \textstyle{{ Charlie}}:\textstyle{{ Bob }}|\textstyle{%
{ Alice}}\right)
\end{array}
\label{eq09}
\end{equation}

Suppose $M=2^{lR_1}$, $N=2^{lR_2}$. Denote

\[
\begin{array}{lll}
\rho _{code} & = & \sum\limits_{aa^{^{\prime }}}\frac 1{MN}\left|
S_{aa^{^{\prime }}}\right\rangle \left\langle S_{aa^{^{\prime }}}\right| \\
\rho _{code}^a & = & \sum\limits_{a^{^{\prime }}}\frac 1N\left|
S_{aa^{^{\prime }}}\right\rangle \left\langle S_{aa^{^{\prime }}}\right| \\
\rho _{code}^{a^{\prime }} & = & \sum\limits_a\frac 1M\left| S_{aa^{^{\prime
}}}\right\rangle \left\langle S_{aa^{^{\prime }}}\right|
\end{array}
\]
and their {\em Von Neumann} entropies by $H_{code}$, $H_{code}^a$, $%
H_{code}^{a^{\prime }}$. According to {\em Holevo} theorem$^{\cite{asH}}$,
the mutual information is bounded by {\em Von Neumann} entropies:

\begin{equation}
\begin{array}{cll}
I\left( \textstyle{{ Charlie}}:\textstyle{{ Alice }},\textstyle{{
Bob}}\right) &
\leq & H_{code} \\
I\left( \textstyle{{ Charlie}}:\textstyle{{ Alice }}|\textstyle{{
Bob}}\right) &
\leq & \sum\limits_{a^{^{\prime }}} \frac 1NH_{code}^{a^{\prime }} \\
I\left( \textstyle{{ Charlie}}:\textstyle{{ Bob }}|\textstyle{{
Alice}}\right) & \leq & \sum\limits_{a} \frac 1MH_{code}^a
\end{array}
\label{eq10}
\end{equation}

Let ${\cal E}$ be the ensemble of letter states that appear as first letters
in the code words, we then have a product distribution, we can define an
entropy $H^{\left( 1\right) }$ and the conditional entropy $H_A^{\left(
1\right) }$ and $H_B^{\left( 1\right) }$. Similarly define $H^{\left(
k\right) }$, $H_A^{\left( k\right) }$ and $H_B^{\left( k\right) }$for each
position $k=1,2,\cdots ,l$ in the code words. According to the subadditivity
of {\em Von Neumann} entropy$^{\cite{aW}}$, we have

\begin{equation}
\begin{array}{cll}
H_{code} & \leq & H^{\left( 1\right) }+\cdots +H^{\left( l\right) } \\
\sum\limits_{a^{^{\prime }}} \frac 1NH_{code}^{a^{\prime }} & \leq &
H_A^{\left( 1\right) }+\cdots +H_A^{\left( l\right) } \\
\sum\limits_{a} \frac 1MH_{code}^a & \leq & H_B^{\left( 1\right) }+\cdots
+H_B^{\left( l\right) }
\end{array}
\label{eq11}
\end{equation}

From Eqs.(\ref{eq09},\ref{eq10},\ref{eq11}) combined, we have

\begin{equation}
\begin{array}{cll}
R_1+R_2 & \leq & \frac 1l\left( H^{\left( 1\right) }+\cdots +H^{\left(
l\right) }\right) \\
R_1 & \leq & \frac 1l\left( H_A^{\left( 1\right) }+\cdots +H_A^{\left(
l\right) }\right) \\
R_2 & \leq & \frac 1l\left( H_B^{\left( 1\right) }+\cdots +H_B^{\left(
l\right) }\right)
\end{array}
\label{eq12}
\end{equation}

Denote $H=\frac 1l\sum\limits_{k=1}^{l}H^{\left( k\right) }$, $H_A=\frac 1l%
\sum\limits_{k=1}^{l} H_A^{\left( k\right) }$, $H_B=\frac 1l%
\sum\limits_{k=1}^{l} H_B^{\left( k\right) }$. Because $\left( H^{\left(
k\right) }-H_B^{\left( k\right) },H_B^{\left( k\right) }\right) $, $\left(
H_A^{\left( k\right) },H^{\left( k\right) }-H_A^{\left( k\right) }\right)
\in {\cal E}$, $k=1,2,\cdots ,l$. According to the convexity of ${\cal E}$,
we know $\left( H-H_B,H_B\right) $ and $\left( H_A,H-H_A\right) $ is also in
${\cal E}$. According to Eq.(\ref{eq03}), all $\left( R_1,R_2\right) $
satisfying Eq.(\ref{eq12}) form a rectangle with an angle cut off, of which $%
\left( H-H_B,H_B\right) $ and $\left( H_A,H-H_A\right) $ are the two outmost
vertices. It follows that it must be $\left( R_1,R_2\right) \in {\cal E}$.
Thus complete the proof of the converse of the theorem.

\section{Some interpretations and applications of the theorem}

The above theorem provides some intriguing quantum communication schemes,
which can be viewed as a generalized superdense coding scheme. The
superdense coding scheme proposed by {\em Bennet} and {\em Wiesner}$^{\cite
{chBsjW}}$ dealt with two-partite communication, but here we will deal with
three-partite communication. Suppose {\em Alice} and {\em Bob} want to send
classical information to {\em Charlie} by $N$-state quantum systems. If {\em %
Alice} and {\em Bob} send message independently, they can send $\log _2N$
bits per system. But we suppose {\em Alice} and {\em Bob} initially share a
considerable supply of $N$-state entanglement, how can they expand their
channel capacity? Note that {\em Alice} and {\em Bob} may be at two distant
locations, so each must encode his/her messages independently of the other
by a predetermined code.

Suppose the initial state {\em Alice} and {\em Bob} shared is $\left| \psi
\right\rangle =\sum\limits_{i=1}^Np_i\left| i\right\rangle _{Alice}\left|
i\right\rangle _{Bob}$, $\rho _0=\left| \psi \right\rangle \left\langle \psi
\right| $, then {\em Alice} and {\em Bob}'s part of the density matrix is $%
\rho _A=tr_B\left( \rho _0\right) $, $\rho _B=tr_A\left( \rho _0\right) $.
Denote $H_E=S\left( \rho ^A\right) =S\left( \rho ^B\right) $, which can be
used to measure the entanglement between {\em Alice} and {\em Bob}'s
systems. {\em Alice} and {\em Bob} can perform a unitary transformation on
his/her systems, then they convey the systems to {\em Charlie} respectively.

Denote $\left\{ T_A\right\} $,$\left\{ T_B\right\} $ as {\em Alice} and {\em %
Bob}'s transformations. They correspond to the $\left\{ \alpha \right\} $,$%
\left\{ \beta \right\} $ discussed before.Then $\rho
=\sum\limits_{T_AT_B}p_{T_A}q_{T_B}(T_AT_B\rho _0T_A^{+}T_B^{+})$ is in $N^2$
dimensional space, so $H\left( \rho \right) \leq 2\log _2N$. Next we examine
the conditional entropy. We see $\rho
_{T_A}=\sum\limits_{T_B}q_{T_B}(T_AT_B\rho _0T_B^{+}T_A^{+})$, because $T_B$
is {\em Bob}'s part of transformation, we have $tr_B\left( \rho
_{T_A}\right) =T_A\rho _AT_A^{+}$, and $H\left( T_A\rho _AT_A^{+}\right)
=H\left( \rho _A\right) =H_E$. According to subaddivity of {\em Von Neumann }%
entropy, we have $H\left( \rho _{T_A}\right) \leq H_E+H\left( tr_A\left(
\rho _{T_A}\right) \right) \leq H_E+\log _2N$. We can similarly have $%
H\left( \rho _{TB}\right) \leq H_E+\log _2N$.

So if Alice and Bob can send information at a rate $\left( R_1,R_2\right) $,
then according to our theorem, it must be

\[
R_1+R_2\leq 2\log _2N
\]

\begin{equation}
\begin{array}{cll}
R_1 & \leq & \log _2N+H_E \\
R_2 & \leq & \log _2N+H_E
\end{array}
\label{eq13}
\end{equation}

Note that all $\left( R_1,R_2\right) $ satisfying Eq.(\ref{eq13}) can be
achieved with {\em Alice} and {\em Bob}'s ensembles of transformation
including all permutations of {\em Schmidt} basis states of the initial
state $\left| \psi \right\rangle $, rotations of the relative phases of
these states, and the combination of the two cases (all with equal
probability).

Although the total amount of information does not increase in the scheme, it
is useful. Because the amount of information {\em Alice} and {\em Bob} want
to send to {\em Charlie} may be different, if {\em Alice} has more
information than {\em Bob} to send, we can adopt a code that increases {\em %
Alice}'s channel capacity at the sacrifice of {\em Bob}'s. This then allows
us to distribute the channel capacity between two users properly, without
the waste of entanglement. From Eq.(\ref{eq13}) we see {\em Alice} can send
information at the maximum rate of $\left( \log _2N+H_E\right) $ bits per
system. In this case, our scheme reduced to the two-partite superdense
coding between {\em Alice} and {\em Charlie}, while {\em Bob} can still send
information to {\em Charlie} at the rate of $\left( \log _2N-H_E\right) $.
This scheme can also be generalized to the case that {\em Alice}, {\em Bob }%
and {\em Charlie} share three-partite entanglement. In this case they can
further expand their channel capacity.

\section{Acknowledgement}

We thank the referee and A.Winter for pointing out to us references\cite{aea}%
\cite{aw}\cite{awi}.

\end{document}